\documentclass[conference]{IEEEtran}
\IEEEoverridecommandlockouts
\usepackage{cite}
\usepackage{amsmath,amssymb,amsfonts}
\usepackage{algorithmic}
\usepackage{graphicx}
\usepackage{textcomp}
\usepackage{gensymb}
\usepackage{xcolor}
\def\BibTeX{{\rm B\kern-.05em{\sc i\kern-.025em b}\kern-.08em
    T\kern-.1667em\lower.7ex\hbox{E}\kern-.125emX}}
\begin{document}

\title{A Workbench for Testing and Simulation Faults in Three-phase Electric Motors with Intelligent Electronic Device and Microcontrolled System\\
}

\author{\IEEEauthorblockN{Giovanni Faria, Michel Fernandes Peres, \\ Osmar Moreira da Silva Neto}
\IEEEauthorblockA{\textit{Electrical Engineering Undergraduate Course} \\ 
\textit{University Center of Arauc\'{a}ria (UNIFACEAR)$ ^1 $}\\
Arauc\'{a}ria, Paran\'{a}, Brazil \\
giovanni.faria@hotmail.com, michelfperes@hotmail.com,\\ osmar\_neto2010@hotmail.com}
\and
\IEEEauthorblockN{Jefferson Rodrigo Schuertz, Edson Leonardo dos \\ Santos and Carlos Alexandre Gouvea da Silva$ ^1 $}
\IEEEauthorblockA{\textit{Electrical Engineering Department (DELT)} \\
\textit{Federal University of Paran\'{a} (UFPR)}\\
Curitiba, Paran\'{a}, Brazil \\
jeffersonschuertz.eng@gmail.com, \\ edson\_l@hotmail.com, carlos.gouvea@ieee.org}
}


\maketitle

\begin{abstract}
Electric motors can be damaged or operate improperly from a possible set of failures. Such failures are related to high or very low voltage and current levels, phase loss or blocked rotor. Therefore, it is important to protect these equipments through appropriate mechanisms.	
Alternatively, a workbench can simulate detectable failures related to the engines, allowing to change parameters, in which maintenance operators are able to identify the results of these changes.
This work presents the development of a workbench as a tool for testing electrical machines and drives. 
The workbench is based on the Arduino programming platform (microcontroller system), in which it checks the functioning of electric motors under the condition of failures that may occur in this engine. 
Motor protections are carried out through an Intelligent Electronic Device (IED), which are popularly known as intelligent relays. The results show the development of a workbench that can test and identify several faults in a small three-phase motor.

\end{abstract}


\section{Introduction}
Electric motors contribute significantly to technological advances in industries and encompass a large part of the energy consumption of countries. 
In Brazil, according to data presented by the Brazilian National Electric Energy Agency (ANEEL) at 2015, the industry consumed 43.7\% of all national electricity and 30\% of all electric energy in the country is consumed by electric motors.
Industrial motor uses a major fraction of the total industrial energy used around the world~\cite{Saidur:2010}.

Electric motors are defined as devices that perform electromechanical energy conversion \cite{Gedra:2004}. 
This electromechanical conversion is explained based on the principles of electromagnetism, in which conductors located in a magnetic field and crossed by electric current are based on the action of a force called torque. 
Electric motors are subdivided into two broad groups: the direct current (DC) and the alternating current (AC)~\cite{Finch:2008}. 
Alternating current motors can be classified as induction or synchronous. The three-phase induction motors are powered by three wires (phases) in which the voltages are 120$^{\circ}$ out of phase~\cite{Neves:2006}.

The basic construction of an alternating current motor is formed by a stationary part, or stator and another rotating part, connected to the axis that couples the motor to load. 
Stator is briefly a metal ring with slits that wrap the coils in a steel core, in which the alternating current flows through the coil producing a rotating magnetic field.
The rotor, on the other hand, is a rod where at its core, there are several conductive bars with uniformly distributed spaces.
When in operation, the rotor core interacts with the magnetic field produced by the stator coils, causing the motor to rotate and generate torque. 
These components are placed inside a housing to protect the engine and control the heat, as showed in Fig. 1.

\begin{figure}[!ht]
	\centering 
	\includegraphics[width=4.5cm]{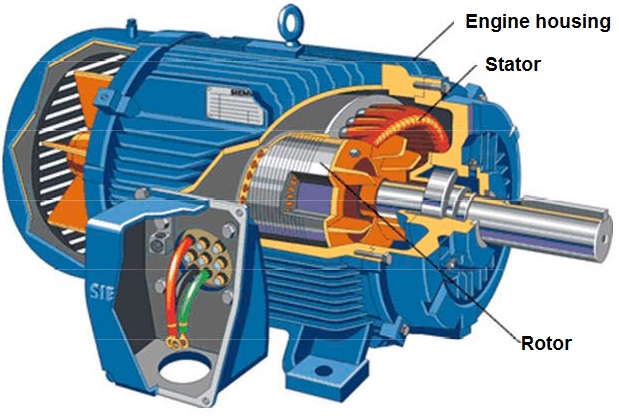} 
	\caption{Three-phase motor.}
\end{figure}

Usually, most of the entire circuit must be protected by internal temperature-sensitive and external devices that interrupt the current, when crossed by a short circuit current.
For the correct protection of the motors, relays are used, which are defined as a wide range of devices that offer protection to electrical systems in many different ways: overload, short circuit, overvoltage, undervoltage and others~\cite{Lee:2002}.
Currently, there is a wide variety of models manufactured from motor protection relays, one of which is the digital relay or Intelligent Electronic Devices (IED).

In the beginning, the main applications of the first IEDs were the protection of transformers, overhead lines and cables, motors and generators. However, it has also been used in protection systems with high reliability as in substations \cite{Sezi:2000}.
IEDs work by means of transformers, such as current transformer (TC) and potential transformer (TP), in which upon receiving an induction of the electric current that flows through the power cable of the motor, it converts high currents and voltages into lower levels.
Thus, the IED receives this information and operates according a programming logic. 
Besides to the circuit protections, the relays also perform functions of communication, control, electrical measurements, signaling and others.
However, test workbenches for these engine protection devices still lack on the market.

Therefore, the development of a workbench for protection of three-phase motors using IED is presented in this paper.
The solution is based on a real-time microcontrolled system that analyzes the behavior of the IED, when typical failures occur in this type of engine.
Besides that, the workbench has a set of commands, so that users can reproduce these failures.

\section{Protection Techniques and Related Works}
Overcurrent is the most common cause of failure in an electrical system, causing a higher level of wear, which reduces the useful lifetime of the components used in the motor supply. 
The overcurrent can be classified as overloads or short circuits. 
Overloads are softened variations in the current flowing in the electrical system. Short circuits, on the other hand, are extreme variations in current flowing in the electrical system.
However, it is challenging to detect and locate these types of failures and interrupting it as quickly as possible~\cite{Baran:2006}.
For induction motors, Zhang \textit{et al.}~\cite{Zhang:2011} suggest investigate five areas of faults for protection methods: thermal protection and temperature estimation, stator insulation monitoring and fault detection, bearing fault detection, broken rotor bar/end-ring detection, and air gap eccentricity detection.

The IEEE Std C37.2$\texttrademark$-2008 \cite{IEEE:2008} is used as a set of function codes and acronyms for equipments used in the equipment installations, generating plants, power substations and power conversion. 
This standard also includes a set of functions used to electric motors protection~\cite{Hunt:2012}. 
The protection relays required to keep AC motors running are based on type, size and application. 
These relays are identified and summarized in IEEE Std C37.96$\texttrademark$-2012 \cite{IEEE:2013}. 

In general, work related to the creation and development of test workbenches for electric motors is limited to the use of applications in educational environments, improving teaching in engineering courses \cite{Silva:2018}.
In these cases, protection simulator designs or the use of electric motors are used by~\cite{Sachdev:1996}\cite{Nigim:2001}\cite{Ernest:2006}\cite{Jothimuni:2016}.

\section{Proposed System and Results}
Fig. 2 shows the main parts of the proposed workbench.
It is used a three-phase induction motor of 220 Volts and 1~HP (horse power) in order to test and verify the accuracy of the proposed system.
At the motor, a blocking system was built and coupled to the axis in order to verify the blocked rotor test.
This mechanism was manufactured from an electromechanical brake and, when energized, interrupts the rotation of the movable motor axis.
The operating conditions of the motor are monitored in real time by the IED.
In this work, an IED manufactured by Schneider using the LTMR 100PBD Tesys model was employed.
This IED is an equipment for monitoring and control motors, in which can provide the following type of protection: load fluctuation, reverse polarity protection, locked rotor, thermal protection, overload (long time), power factor variation, earth-leakage protection, thermal overload protection, phase unbalance, and phase loss.

\begin{figure}[!ht]
	\centering 
	\includegraphics[trim = 5mm 5mm 5mm 5mm, clip, width=6cm]{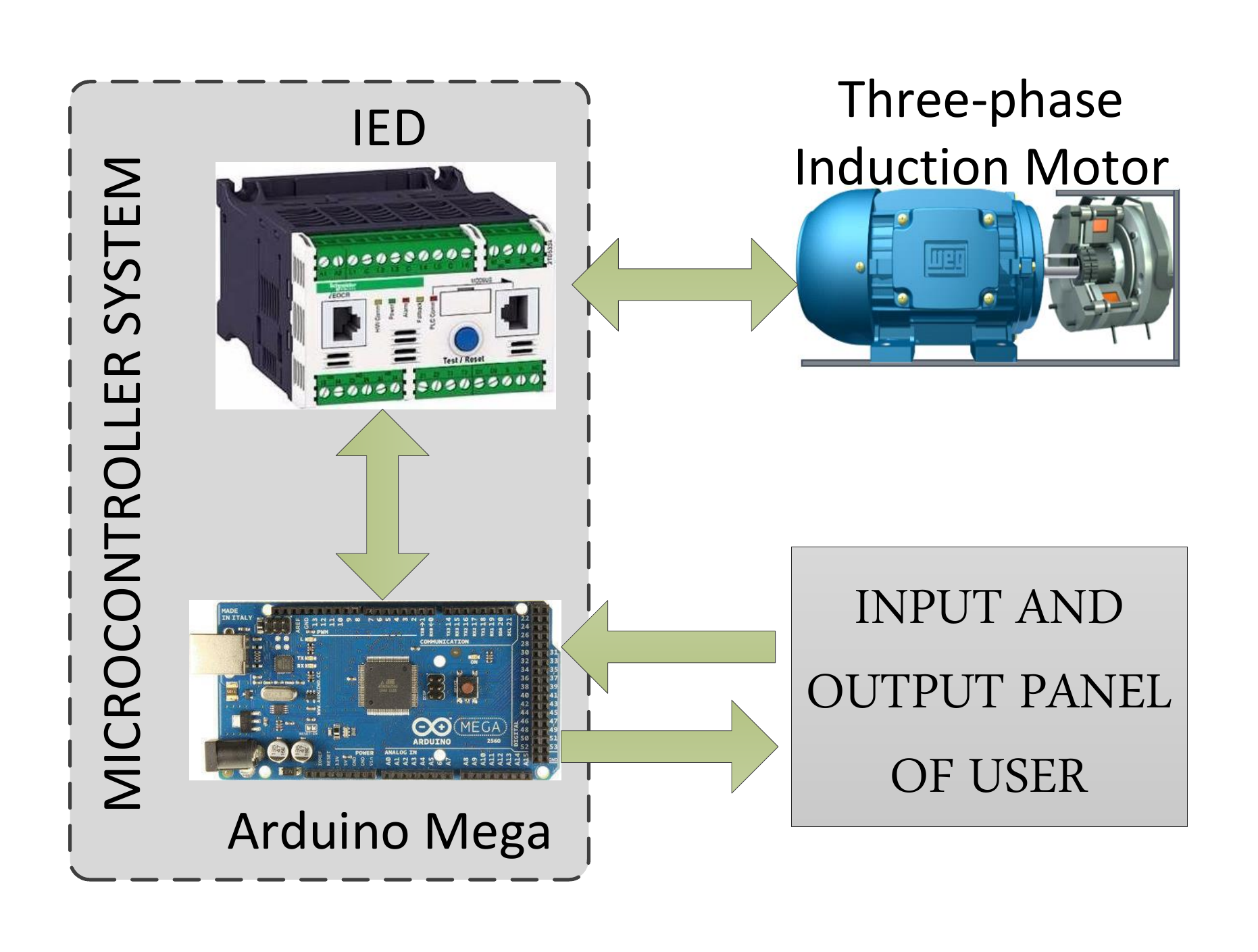} 
	\caption{Workbench for Protection Motor Diagram.}
\end{figure}

From each type of failure that the IED can identify and protect the motor, it sends a notification message to Arduino.
This message is identified by the microcontroller, which in turn sends a response to the fault identified by the IED to the user panel.
This response can be of three types: a message written on a liquid crystal display (LCD), on indicative LEDs or an audible signal from a buzzer.

In order to develop and implement the workbench, we focus the main types of failures that occur in three-phase motors.
The main types of faults are indicated by: (1) Overvoltage; (2)~Undervoltage; (3) Overcurrent; (4) Phase loss; (5) Locked rotor; (6) Extended start; (7) Voltage unbalance; and (8) Current unbalance. 
The motor protection used in the workbench was based on the specifications available in the IEEE C37.2$\texttrademark$-2008 e IEEE C37.96$\texttrademark$-2012 standards.
To simulate the failures are used input elements such as on/off keys, buttons and a potentiometer that allow increase or decrease voltage at the motor.
In order to simulate a blocked rotor failure, a key is used to block the motor rotation system.

\subsection{Microcontroller System}
The programming of the intelligent relay (IED) was developed in the software that accompanies the relay called Tesys~T, which uses block diagram language. In this programming, values were set that parameterize the relay, where for each failure a value is pre-established.

The criteria configured via software at IED for fault detection are shown as follows:

\begin{itemize}
	\item For the overcurrent fault to occur, the current must vary between 20\% to 800\% of the rated current value. For the fault operation in the project, the value of 120\% of the nominal current was parameterized. 
	\item In the case of blocked rotor failure, the same current variation occurs, however the value set in the parameterization was 130\% of the nominal current. In the current unbalance fault, this interval varies between 10\% to 70\% the value of the percentage of unbalance, where the value defined in the project to occur the fault was 10\%.
	\item In the extended start, the variation occurs between 100\% to 800\% of the nominal current value, where the value of 150\% was defined for the failure action in the project.
	\item In the case of undervoltage failure, the ideal interval should occur if the nominal voltage is greater than or equal to 70\% and less than or equal to 99\%. In the case of overvoltage, this range varies from 101\% to 115\%. For parameterization purposes, the value of 85\% for undervoltage and 110\% for overvoltage was selected.
	\item In the case of voltage unbalance failure, the voltage analyzed is not the nominal voltage, but the motor phase voltage, where the variation is at least 3\% and at most 10\%, with the value being 10\%.
\end{itemize}

The program was developed in C language and built-in Arduino microcontroller to receive the IED signals, as well as send messages and failure states on the user panel. This panel is illustrated in Fig. 3.

\begin{figure}[!ht]
	\centering
	\label{fig:comandos}
	\includegraphics[trim = 48mm 57mm 59mm 33mm,clip,width=9cm]{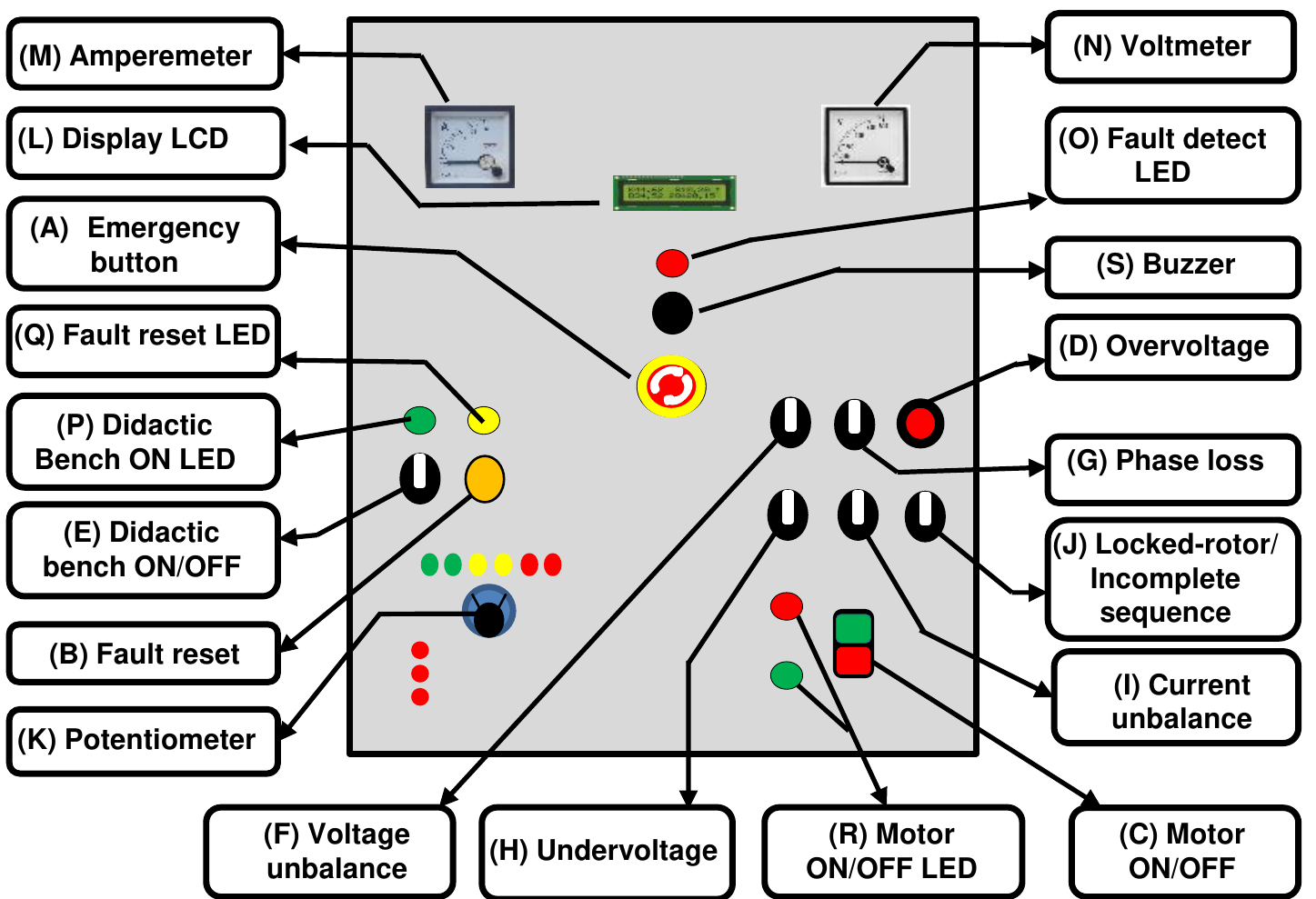}
	\caption{Control panel of the workbench.}
\end{figure} 


Fig. 4 shows the engine with the electromechanical brake system.
For safety reasons, this motor is protected in a housing with a safety sensor. If the engine is running and the case is opened, the system will shut off the engine's power supply.

\begin{figure}[!ht]
	\centering
	\includegraphics[width=6cm]{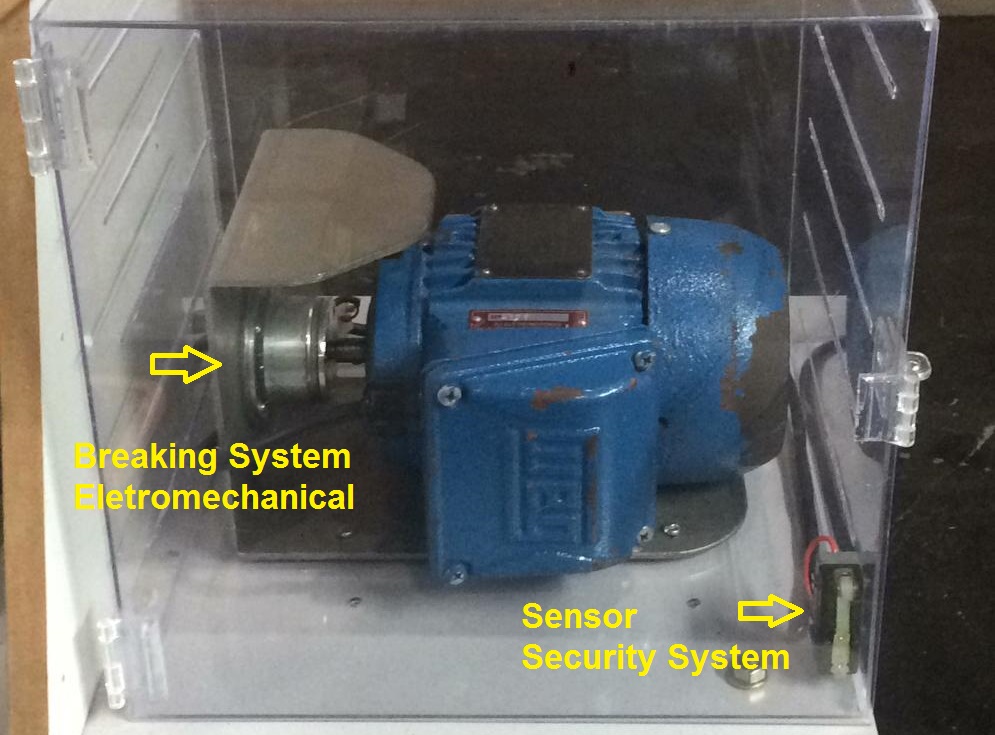}
	\caption{The overview of motor and break system.}
\end{figure}

Fig. 5 shows the final result of the workbench construction. 
From the transparent material on the front of the workbench, users can check the main internal components and their connections, which can prompt and encourage users to ask questions related to the protection of the motor and its devices.

\begin{figure}[!ht]
	\centering
	\includegraphics[width=6cm]{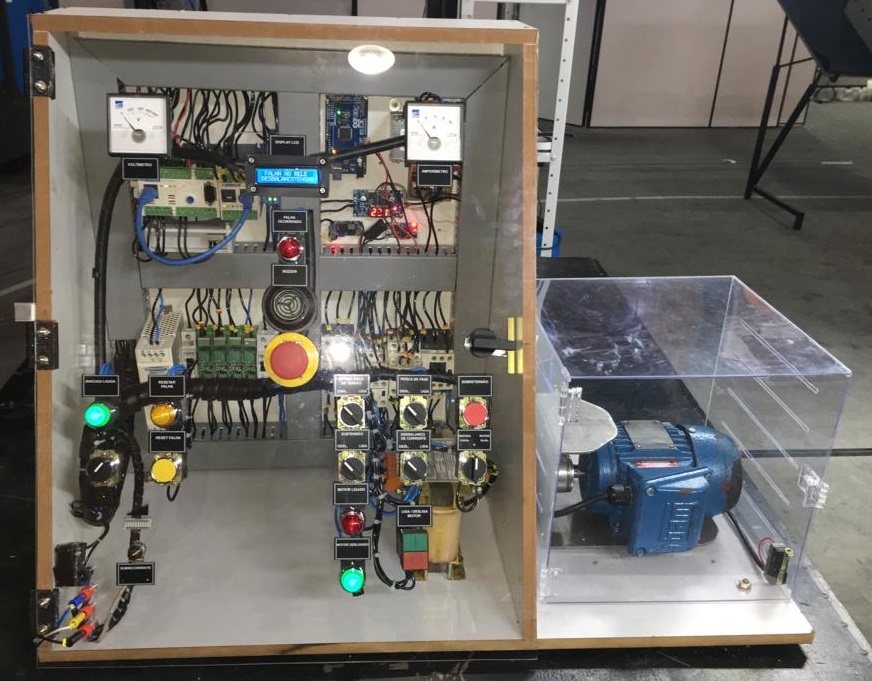}
	\caption{The overview of full workbench.}
\end{figure}

\begin{figure}[!ht]
	\centering
	\label{fig:ligada}
	\includegraphics[width=6cm]{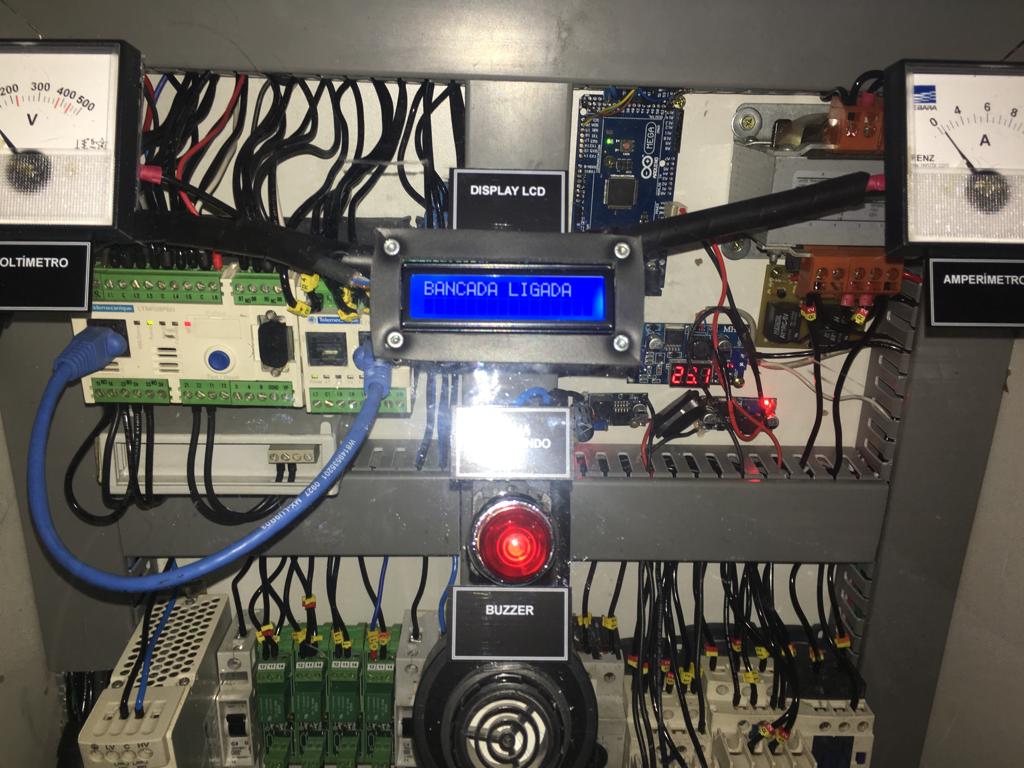}
	\caption{LCD with workbench message ON.}
\end{figure}

\subsection{Test Procedures}
First, to start the fault simulation, it is necessary to turn on the workbench switch, where the green LED will light.
In addition, the LCD displays a ``Workbench Working'' message, as shown in Figure 6.
Next, it should be shown that the IED is operational in order to identify the faults that are sent by the user.
To create a motor fault, the yellow fault reset LED cannot be lit.
If the LED is on, the IED must be reset using the reset button failure.

\subsubsection{Overvoltage Failure}
Overvoltage occurs when there is a voltage above the nominal motor supply.
To perform the overvoltage failure, a three-phase transformer (II) was connected to one of the contactors (8).
When the overvoltage button (D), this contactor coil is activated, allowing the transformer to send 380~Volts to the motor.
Thus, when the IED identifies that there is a voltage above the nominal motor voltage (220~Volts), it triggers the motor trip, sending a digital signal to the Arduino to indicate on the display (L) the occurrence of the failure.



\subsubsection{Undervoltage Failure}
To simulate this fault, it is necessary that the three-phases of the motor supply are interrupted abruptly.
In this way, the drive relays were interconnected to the power contactors, so that when the undervoltage switch is activated, the coil of the three relays is powered, making it impossible to supply voltage to the motor.

\subsubsection{Overcurrent Failure}
To simulate this load, a braking system was connected to the motor axis; therefore, when the motor is already running and the potentiometer is turned slightly to the point of overcurrent event, the voltage will decrease at the magnetic brake.
Then, lightly locking the motor axis so that it is not completely locked.

\subsubsection{Phase Loss Failure}
When the phase failure selector switch is activated, the coil of the two command relay is powered, making it impossible for the three-phases to reach the motor.

\subsubsection{Locked Rotor Failure}
To simulate this failure, the magnetic brake was used again, but this time this brake will be activated through the locked rotor selector. When moving the key to the right side, the motor axis will be completely locked, as the voltage is 0 Volts.

\subsubsection{Extended Starting Failure}
For complete blocking of the motor, the same selector switch is used that generates the failure of the blocked rotor; however, for extended starting failure, the switch must be moved to the left.
Thus, the IED recognizes that the motor is stopped and after programmed timeout in the IED, that the motor does not start is disarmed. 

\subsubsection{Voltage Unbalance Failure}
To simulate this failure, the drive relays were interconnected to the power contactors, so that when the voltage unbalance selector switch is activated, a relay coil is powered, making it impossible for the relay voltage phase to reach the motor.

\subsubsection{Current Unbalance Failure}
In order to simulate the driver relays, they were interconnected with the power contactors, so that when the current unbalance selector switch is activated, a relay coil is powered, making it impossible for the voltage of this phase to reach the motor.

For each type of fault, 30 repetitions were performed in total to check the reliability of the system in identifying each type of fault and in protecting the motor.
Only two types of failure did not reach full success: undervoltage with 70\% and phase loss with 86.6\%.
This is due to the fact that there is a great similarity between the two types of failure, where sometimes the EDI interpreted that the undervoltage occurs and other times it identified that the failure occurred was phase loss.

Considering that this test workbench can be used as a teaching tool, it is possible to make a comparison with~\cite{Silva:2018}. The main advantage of the work proposed by~\cite{Silva:2018} is the low cost due to the use of virtual simulators or emulators only. Otherwise, our work can be used in real practical classes and not just simulated. In addition, it also allows the study of real equipments, unlike the use of only simulators. 

This workbench is limited to the use of three-phase motors based on the nominal motor voltage and also does not provide support for DC motors.
For motors with a high power level, it is necessary to update the IED firmware.


\section{Conclusion}
In this work, a test workbench and fault simulation in induction motors was proposed.
System control is based on the Arduino programming platform, whereas motor protection is performed by an Intelligent Electronic Device.
In a control panel, it was possible to verify the performance of the IED in eight types of common faults in the motor under test.
The results showed that the workbench was able to identify most of the failures implemented with a high level of confidence.
As a future work, it is suggested to use this workbench as an educational teaching tool in electrical engineering courses.


\begin{thebibliography}{00}
	
\bibitem{Saidur:2010} R. Saidur, ``A review on electrical motors energy use and energy savings,'' \textit{Renewable and Sustainable Energy Reviews}, vol. 14, no. 3, pp. 877--898, 2010.	

\bibitem{Gedra:2004} T. W. Gedra, S. An, Q. H. Arsalan and S. Ray, ``Unified power engineering laboratory for electromechanical energy conversion, power electronics, and power systems,'' \textit{IEEE Transactions on Power Systems}, vol. 19, no. 1, pp. 112--119, 2004.

\bibitem{Finch:2008} J. W. Finch and D. Giaouris, ``Controlled AC Electrical Drives,'' \textit{IEEE Transactions on Industrial Electronics}, vol. 55, no. 2, pp. 481--491, 2008.

\bibitem{Neves:2006} P. N. C. Neves and J. L. Afonso, ``Traction system for electric vehicles using a variable frequency three-phase induction motor driver with regenerative braking,'' in \textit{3rd International Conference on Hands-on-Science - Science Education and Sustainable Development}, Braga, Porgutal, 2006.

\bibitem{Lee:2002} W.-J. Lee, J.-C. Gu, R.-J. Li and P. Didsayabutra, ``A physical laboratory for protective relay education,'' \textit{IEEE Transactions on Education}, vol. 45, no. 2, pp. 182--186, 2002.

\bibitem{Sezi:2000} T. Sezi and B. K. Duncan, ``New intelligent electronic devices change the structure of power distribution systems,'' in \textit{Conference Record of the 1999 IEEE Industry Applications Conference. Thirty-Forth IAS Annual Meeting}, Phoenix, AZ, USA, pp. 944--952, 1999.

\bibitem{Baran:2006} M. E. Baran and N. R. Mahajan, ``Overcurrent Protection on Voltage-Source-Converter-Based Multiterminal DC Distribution Systems,'' \textit{IEEE Transactions on Power Delivery}, vol. 22, no. 1, pp. 406--412, 2007.

\bibitem{Zhang:2011} P. Zhang, Y. Du, T. G. Habetler and B. Lu, ``A Survey of Condition Monitoring and Protection Methods for Medium-Voltage Induction Motors,'' \textit{IEEE Transactions on Industry Applications}, vol. 47, no. 1, pp. 34--46, 2011.

\bibitem{IEEE:2008} IEEE, ``IEEE Standard Electrical Power System Device Function Numbers, Acronyms, and Contact Designations,'' in ``IEEE Std C37.2-2008'' (Revision of IEEE Std C37.2-1996) , pp.1-48, 3 Oct. 2008.

\bibitem{Hunt:2012} R. Hunt, R. Luna and S. Patel, ``Protecting Remotely Located Motors: Application Issues and Solutions to the Technical Challenges,'' \textit{IEEE Industry Applications Magazine}, vol. 18, no. 6, pp. 37--49, 2012.

\bibitem{IEEE:2013} IEEE, ``IEEE Draft Guide for AC Motor Protection,'' in IEEE PC37.96/D11, pp.1--141, 23 Aug, 2012.

\bibitem{Silva:2018} C. A. G. Silva, E. L. Santos and D. A. F. Pelacini,``Evaluation of Academic Experience in Learning Education over Simulators Softwares,'' \textit{International Journal on Alive Engineering Education}, vol. 5, no. 2, pp. 23--40, 2018.

\bibitem{Sachdev:1996} M. S. Sachdev and T. S. Sidhu, ``A laboratory for research and teaching of microprocessor-based power system protection,'' \textit{IEEE Transactions on Power Systems}, vol. 11, no. 2, pp. 613--619, 1996.

\bibitem{Nigim:2001} K. A. Nigim and R. R. DeLyser, ``Using MathCad in understanding the induction motor characteristics,'' \textit{IEEE Transactions on Education}, vol. 44, no. 2, pp. 165--169, 2001.

\bibitem{Ernest:2006} E. Ernest, R. Sztylka, B. Ufnalski and W. Koczara, ``Methods in Teaching Modern AC Drives: Inverter-fed Motor System with Internet-based Remote Control Panel,'' \textit{2006 12th International Power Electronics and Motion Control Conference}, Portoroz, pp. 2130--2133, 2006.
	
\bibitem{Jothimuni:2016} R. S. M. Jothimuni, H. M. D. M. B. Wijerathne, C. A. N. Yapa, D. W. N. T. Wijethunga, J. R. Lucas and P. S. N. de Silva, ``Power System Simulator - a teaching tool protection integration,'' in \textit{2016 Moratuwa Engineering Research Conference}, Moratuwa, pp. 231--236, 2016.	
	
	

%
%

\end{thebibliography}
\end{document}